\documentclass[twocolumn]{aastex701}
\usepackage{amsmath,amstext}
\usepackage[T1]{fontenc}
\usepackage{apjfonts} 
\usepackage{multirow}
\usepackage[figure,figure*]{hypcap}
\usepackage{booktabs,caption}
\usepackage{tablefootnote}
\usepackage{hyperref}
\usepackage{subcaption}
\usepackage{mwe}

\begin{document}

\title{The AstroSat UV Deep Field South-V: Constraints on the average escape \\ of ionizing photons in the cosmic dusk}

\author[0009-0003-8568-4850]{Soumil Maulick}
\affil{Inter-University Centre for Astronomy and Astrophysics, Ganeshkhind, Post Bag 4, Pune 411007, India}
\email[show]{soumil@iucaa.in} 

\author[0000-0002-8768-9298]{Kanak Saha}
\affil{Inter-University Centre for Astronomy and Astrophysics, Ganeshkhind, Post Bag 4, Pune 411007, India}
\email[show]{kanak@iucaa.in}



\begin{abstract}
We investigate the escape of ionizing (Lyman-continuum; LyC) photons from 49 star-forming galaxies at redshifts $\sim 1-1.5$, using far-ultraviolet (FUV) imaging from the Ultra-Violet Imaging Telescope (UVIT) onboard AstroSat. The sample spans a wide range of stellar masses (${8.3\lesssim\text{log}_{10}(\text{M}_{*}/\text{M}_{\odot})\lesssim10.1}$) and UV luminosities (${-20.6 \lesssim }{M}_{\text{UV}}\lesssim-{17.5}$). LyC emission is undetected in most galaxies (42/49), and stacking these galaxies yields only an upper limit on the observed LyC-to-nonionizing UV flux density ratio ($({F}_{\lambda,\rm{LyC}}/{F}_{\lambda,\rm{UV}})_{\rm{obs}}<0.12$). {This corresponds to an absolute escape fraction upper limit of a few percent} (${\langle f_{\text{esc,abs}} \rangle < 0.03}$), {under a range of stellar population assumptions}. Including all galaxies (with 7 LyC-leaker candidates) produces a marginal $2.4\sigma$ detection, suggesting that the average LyC signal is driven by a small number of sources. 

To identify the conditions favorable for LyC escape, we perform stacking analyses in bins of stellar mass, UV slope, compactness, inclination, and star formation rate surface density. A stacked LyC signal is detected at a significance of $\sim3\sigma$ from a subset of galaxies that are characterized as being compact, have high star formation rate surface densities, and blue UV continuum slopes, despite each of these being individually undetected in LyC. This provides the first systematic evidence at $z\sim1-1.5$ linking these properties to LyC escape, consistent with trends observed in the lower redshift universe. Additionally, LyC leakage appears more efficient in low-mass galaxies ($\log_{10}(\text{M}_{*}/\text{M}_{\odot})<9.5$), with their average absolute escape fraction ranging from $\langle f_{\text{esc,abs}} \rangle\sim0.1-0.2$ depending on stellar population assumptions. These results support the scenario that compact, low-mass starbursts were key contributors to the ionizing photon budget during cosmic reionization.  
\end{abstract}



\section{Introduction}
The Epoch of Reionization (EoR) marks a major phase transition in cosmic history, during which neutral hydrogen in the intergalactic medium (IGM) was reionized. Observational constraints on its timing include the Gunn–Peterson effect \citep{Gunn65,Becker01} and Ly$\alpha$ optical depth measurements \citep{Bosman22}, which indicate that hydrogen reionization was largely complete by $z \sim 5.3$. However, the nature of the sources that are responsible for reionization remains uncertain. Star-forming galaxies, particularly low-mass galaxies, are widely considered to drive this process \citep{Finkelstein19,Atek24}, although alternative models suggest that more massive galaxies \citep{Naidu20} or even active galactic nuclei (AGN; \citealt{Madau24,Dayal25}) may contribute significantly. Empirical progress has been made in constraining two key components of the ionizing photon budget during the EoR. The first is the UV luminosity function of galaxies, which upon integration yields the cosmic UV luminosity density. The second is the intrinsic production rate of ionizing photons. Both components have been significantly informed by recent JWST observations \citep{Simmonds23,Stark25}. The third and most uncertain component is the fraction of LyC photons that escape their host galaxies into the IGM ($f_{\text{esc}}$). At $z \gtrsim 4$, direct observations of LyC emission are strongly hindered by the rising IGM opacity \citep{Inoue14}, making lower-redshift galaxies that exhibit LyC leakage valuable analogs for understanding sources of reionization. 

The past decade has witnessed substantial observational progress in identifying galaxies that leak Lyman-continuum (LyC) photons, known as LyC leakers, across a wide range of cosmic epochs. Even well-known local LyC leakers discovered earlier, such as Haro 11 \citep{Bergvall00,Bergvall06}, have been revisited with high-resolution imaging and spectroscopy in recent years \citep{Menacho19,Ostlin21,LeReste23,Komarova24}, yielding new insights into the mechanisms of LyC escape in nearby analogs.

In the redshift range $z\sim0.2$–0.4, the Low-redshift Lyman Continuum Survey (LzLCS; \citealt{Flury22a}) has been particularly influential. The combined LzLCS+ sample, consisting of 66 galaxies from LzLCS and 23 from prior studies \citep{Izotov16b,Izotov18,Izotov18b,Izotov21,Wang19}, exhibits a high LyC detection rate ($\gtrsim50\%$) with HST–COS, highlighting the effectiveness of the target-selection strategy. Statistical analyses that incorporates censored data of the LzLCS+ sample \citep{Jaskot24,Jaskot24b} further identify key observational predictors of LyC leakage, including the observed UV continuum slope ($\beta_{\rm obs}$, where $F_{\lambda}\propto\lambda^{\beta}$; \citealt{Calzetti94}), the $[\mathrm{O\: III}]\lambda5007/[\mathrm{O\:II}]\lambda3727$ (O32) ratio, and elevated star-formation rate surface densities ($\Sigma_{\rm SFR}$). These trends suggest that LyC escape is more likely in galaxies hosting young, vigorously star-forming stellar populations, consistent with theoretical expectations that feedback-driven processes can carve low-opacity channels through the ISM \citep{Kimm14,Trebitsch17,Ma20,Rosdahl22,Carr25,Flury25}. Nonetheless, the presence of non-detections and only weak LyC leakage in some galaxies with these same hallmarks highlights the complexity of the phenomenon, pointing to additional influences such as the timing of mechanical feedback relative to star formation, the stochasticity of LyC escape, and anisotropy in the escape channels \citep{Flury25}.

At the other end of the LyC leaker landscape, the Keck Lyman Continuum Spectroscopic Survey (KLCS; \citealt{Steidel18}), a deep spectroscopic survey has probed LyC emission from 136 star-forming galaxies at $z\sim3$. These studies reveal that the majority of galaxies show only modest LyC leakage, with a mean escape fraction of $\langle f_{\rm esc}\rangle \sim 0.06$ \citep{Pahl21}. UV imaging programs, such as UVCANDELS \citep{Wang25} with HST/WFC3, have placed stringent upper limits on average LyC escape through stacking analyses in the redshift range $z\sim2$-4. At these higher redshifts, however, the increasing opacity of the intergalactic medium (IGM) introduces a strong degeneracy with ISM escape, requiring careful modeling to disentangle the two effects \citep{Steidel18,Bassett21}.

At intermediate redshifts ($z \sim 1-1.5$), HST grism surveys such as 3D-HST \citep{Momvheva16} and CLEAR \citep{Simons23} have been used to identify star-forming H$\alpha$ emitters, that are subsequently probed for LyC leakage with far-UV imaging from GALEX \citep{Rutkowski16}, HST ACS/SBC \citep{Siana07,Siana10,Alavi20,Jung24}, and UVIT \citep{Saha20,Dhiwar24,Maulick24,Maulick25}. FUV imaging in this redshift range typically probes LyC photons at bluer wavelengths ($\sim750\: \text{\AA}$) than those sampled in both lower- and higher-redshift studies. At such wavelengths, the effective optical depth to neutral hydrogen absorption is about half that at the Lyman limit ($\sim912\: \text{\AA}$). This intermediate-redshift regime provides a vital bridge for testing whether the LyC leakage trends identified at lower redshifts remain valid at epochs approaching the peak of the cosmic star-formation rate density \citep{MD14}, where the intergalactic medium retains sufficient transparency ($T_{\text{IGM}}\gtrsim0.5$).
GALEX- and HST SBC-based studies report no individual detections of LyC emission, instead placing upper limits on LyC-to-nonionizing UV flux density ratios and escape fractions, both for individual sources and stacked samples. These efforts span a diverse set of galaxy populations, including low-mass star-forming galaxies \citep{Rutkowski16,Alavi20} and galaxies with extremely blue UV continua \citep{Jung24}. 

By contrast, UVIT observations have revealed individual LyC-leaking galaxies at similar redshifts. However, no UVIT study has yet undertaken a systematic investigation of LyC escape. In this work, we carry out such an analysis using F154W imaging from the AstroSat UV Deep Field South (AUDFs) Survey \citep{Saha24}. We investigate LyC leakage in a sample of 49 star-forming galaxies, assembled from publicly available spectroscopic catalogs and complemented with high-resolution imaging (Section \ref{sec:selection}). The properties of the sample are summarized in Section \ref{sec:properties}. Our stacking methodology in the F154W band is detailed in Section \ref{sec:stacking}, while the resulting stacks, including analyses of subsamples defined by galaxy properties associated with LyC escape are presented in Section \ref{sec:results}. We discuss the implications of our findings in Section \ref{sec:discussion} and provide a summary in Section \ref{sec:summary}.

All magnitudes quoted are in the AB system \citep{Oke83}. All flux measurements have been corrected for Galactic extinction using the color excess from \citet{Schlafly11} and the \citet{Cardelli89} extinction curve. Throughout this work, we adopt the concordance $\Lambda$CDM cosmological model, with $H_0=70\: \text{km}\:\text{s}^{-1}\:\text{Mpc}^{-1}$, $\Omega_m =0.3$, $\Omega_{\Lambda}=0.7$. All flux ratios presented correspond to flux densities in the wavelength space ($F_{\lambda}$).

\section{Data}
We make use of FUV imaging (F154W band) of the GOODS-South field \citep{Giavalisco04} from the AstroSat UV Deep Field South survey (AUDFs, \citealt{Saha24}), obtained with UVIT. The depth of the F154W mosaic varies across the field, and for this study we restrict ourselves to the deepest regions (mean exposure $\sim63,200$ s), which achieve a $3\sigma$ limiting magnitude of 27.3 in a circular aperture of radius 1.4$''$.

For ancillary data, we employ high-resolution HST imaging from the Hubble Legacy Fields\footnote{\href{https://archive.stsci.edu/prepds/hlf/}{https://archive.stsci.edu/prepds/hlf/}} (HLF) program \citep{Illingworth16,Whitaker19}, combining observations from WFC3/UVIS, ACS/WFC, and WFC3/IR. We also make use of JWST/NIRCam imaging from the JADES survey \citep{Eisenstein23,Rieke23}\footnote{\href{https://archive.stsci.edu/hlsp/jades}{https://archive.stsci.edu/hlsp/jades}}, primarily for visual verification of sources and for identifying ''empty-sky'' positions (Section \ref{sec:stacking}).

\section{Sample selection} \label{sec:selection}
All our targets lie within the deepest region of the AUDFs which overlaps significantly with the GOODS-South field. This field benefits from extensive multi-wavelength coverage from both ground- and space-based observatories, enabling us to leverage several spectroscopic surveys to identify galaxies with secure spectroscopic redshifts.

Our primary goal is to select galaxies in the redshift range $0.97 < z < 1.55$ that are sufficiently isolated with respect to the point spread function (PSF) of the UVIT F154W band. This isolation criterion is crucial to minimize the risk of foreground nonionizing contamination \citep{Siana07,Vanzella10} when measuring Lyman-continuum (LyC) emission, either from individual galaxies or from stacked ensembles. The lower redshift limit of $z \gtrsim 0.97$ ensures that the red end of the F154W bandpass falls completely blueward of the Lyman limit of the selected sources such that this filter probes pure LyC emission.

To identify suitable targets, we utilize a combination of the following spectroscopic catalogs:
\begin{itemize}
    \item the MUSE Hubble Ultra Deep Field (HUDF) DR2\footnote{\href{https://amused.univ-lyon1.fr}{https://amused.univ-lyon1.fr}} catalog \citep{Bacon23},
    \item the MUSE WIDE Survey\footnote{\href{https://musewide.aip.de/}{https://musewide.aip.de/}} \citep{Herenz17,Urrutia19},
    \item the CLEAR Survey\footnote{\href{https://archive.stsci.edu/hlsp/clear}{https://archive.stsci.edu/hlsp/clear}} \citep{Simons23} and,
    \item the 3D-HST Survey\footnote{\href{https://archive.stsci.edu/prepds/3d-hst/}{https://archive.stsci.edu/prepds/3d-hst/}} \citep{Momvheva16}.
\end{itemize}
For the selected redshift range, the dominant spectral feature in the MUSE catalogs (HUDF and WIDE) is the [O II]$\lambda \lambda\:3727,3729$ emission line, while in the grism-based HST surveys (CLEAR and 3D-HST), it is the H$\alpha$ emission line captured with the G141 grism.

In the first stage of selection, we choose galaxies from the MUSE HUDF DR2, MUSE WIDE, and CLEAR catalogs within the redshift range $0.97 < z < 1.55$. From the MUSE catalogs, we include only those sources with a spectroscopic confidence flag of 3, indicating secure redshifts. For the CLEAR catalog sources, we require a signal-to-noise ratio (S/N) greater than 3 in the H$\alpha$, H$\beta$, and [O III]$\lambda\lambda$ 4959,5008 lines to ensure similar confidence in redshift determination.

The resulting sample is then cross-matched with the Hubble Legacy Field (HLF) photometric catalog \citep{Whitaker19} to identify galaxies with no neighboring sources within a 1.4$''$ radius. This initial isolation criterion is imposed with stacking in mind, trying to ensure that the targets in the F154W band remain uncontaminated by blending with foreground objects.
This initial selection yields 65, 89, and 65 galaxies from the MUSE HUDF, MUSE WIDE, and CLEAR catalogs, respectively. In the next step, we focus on the MUSE-selected galaxies and further downsize this sample by cross-matching them with the CLEAR and 3D-HST surveys. We retain only those galaxies that exhibit H$\alpha$ emission with S/N$ > 3$ in the HST grism catalogs. This additional filtering not only strengthens the reliability of the spectroscopic redshift determination, but is also physically motivated, as we subsequently derive quantities such as the star-formation rate surface density from the H$\alpha$ fluxes.

Following this step, the MUSE HUDF and MUSE WIDE samples are reduced to 47 and 66 galaxies, respectively. We then combine these with the CLEAR sample to obtain a total of 161 unique galaxies, ensuring that duplicates are not double-counted.

Next, we perform a visual inspection of each of these 161 sources.  Only galaxies located within the deepest region of the F154W mosaic, an eye-shaped area defined in \citet{Saha24}, are retained, resulting in the exclusion of 31 galaxies.

From the remaining sample, an additional 69 galaxies are removed due to one or more of the following:
(1) Visual inspection of high-resolution HST (from the HLF \citealt{Illingworth16,Whitaker19}) and/or JWST (from the JADES survey; \citealt{Eisenstein23,Rieke23}) images revealed either nearby contaminants or proximity to bright extended sources that may affect the F154W measurements,
(2) Uncertainty in the spectroscopic redshift or lack of a convincing H$\alpha$ detection upon inspection of the spectra. 

After the manual vetting, the sample comprises 61 galaxies. We classify the sources as LyC-leaker candidates or non-detections based on the presence (or absence) of a counterpart and its significance in the F154W-band image. For each object in our sample, we perform forced aperture photometry in the F154W band using 1.4$''$ radius circular apertures placed at the object's centroid in the HST F606W band, and assess the flux significance using the noise map from \citet{Saha24}. We also take into account the local residual  background in the F154W image, discussed in Section \ref{sec:stacking} while estimating the significance.
Sources with S/N > 2 and having counterparts in both the F154W (probes rest-frame LyC) and N242W (probes rest-frame non-ionizing UV) Source Extractor \citep{Bertin96} generated catalogs in \cite{Saha24} are classified as LyC leaker candidates. The rest, we classify as non-detections.

\begin{figure}[ht!]
\nolinenumbers
\includegraphics[width=0.45\textwidth]{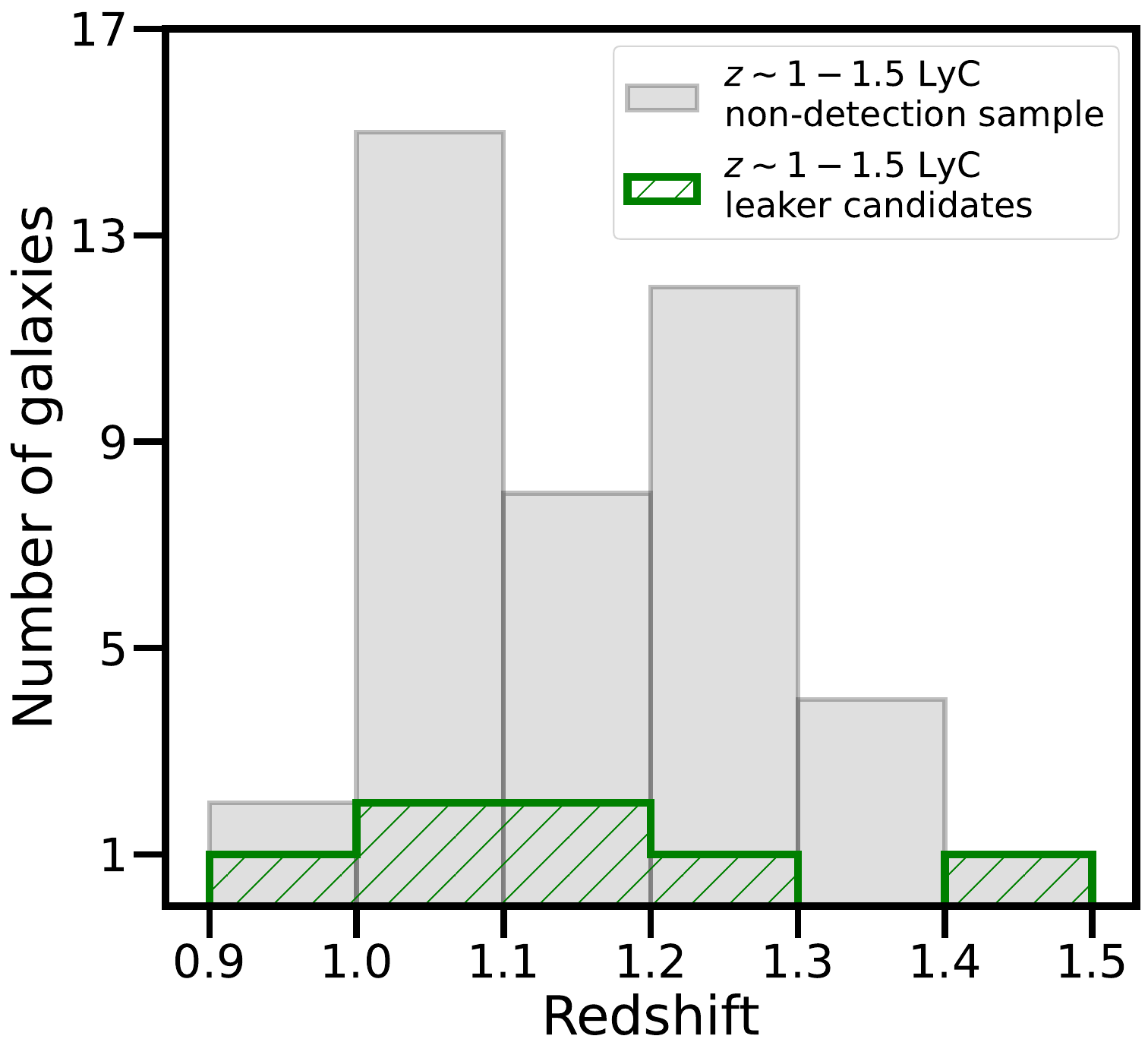} 
\caption{Redshift distribution of our selected sample of 49 star-forming galaxies.} 
\label{fig:z_distbn}
\end{figure}

\begin{figure*}[ht!]
\nolinenumbers
\centering
\includegraphics[width=0.8\textwidth]{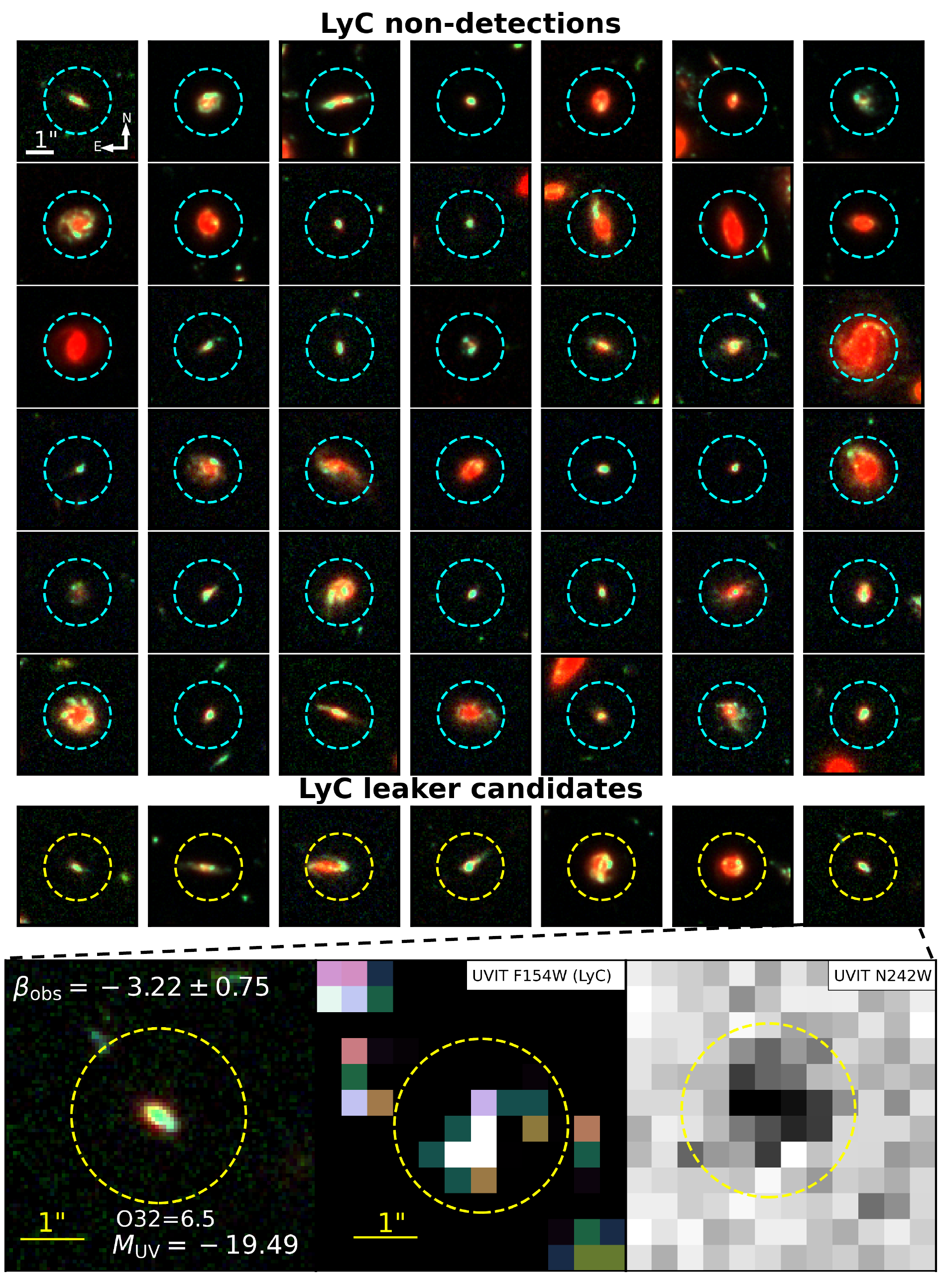}
\caption{RGB composite images of galaxies in our sample, constructed from HST imaging: F160W (red), F606W (green), and F435W (blue). Each image is displayed within a circular aperture of radius $1.4''$. The top six rows show LyC non-detections (cyan dashed apertures), while the bottom row presents the LyC leaker candidates (yellow dashed apertures) as defined in Section \ref{sec:selection}. The final panel shows the last LyC leaker candidate at $z\sim1.23$ (F154W ID: AUDFs\_F15450), newly identified in this study, in an enlarged view, along with its observed UV continuum slope ($\beta_{\text{obs}}$), O32, and absolute UV magnitude ($M_{\text{UV}}$) values. Its UVIT F154W and N242W counterparts are highlighted within the same aperture to the right of the RGB image. Within the yellow dashed aperture (radius 1.4''), the S/N of this object is $\sim2.9$ in the F154W band image. A noise peak is evident toward the northeast in the F154W band. The F154W and N242W bands probe the LyC ($\sim690\:\text{\AA}$) and the rest-frame non-ionizing UV ($\sim1085\:\text{\AA}$) emission, respectively. The F154W band image has been smoothed with a top-hat kernel of radius 1.5 pixels to enhance visual clarity.} 
\label{fig:stamp_images1}
\end{figure*}

This results in 8 LyC leaker candidates and 53 non-detections. Of the 8 candidates, 7 have been previously reported in \citet{Maulick24, Maulick25}, Maulick et al. (2025b, submitted), and we identify one new candidate (F154W ID: AUDFs\_F15450) from the MUSE WIDE sample. Additionally, we include 1 more candidate from \citet{Maulick25} in the LyC leaker candidate sample. In the final step of our selection, we search for X-ray counterparts by cross-matching our sample with the Chandra Deep Field-South 7 Ms catalog \citep{Luo17}. Although some X-ray-detected sources are classified as “Galaxies” in \citet{Luo17}, we conservatively exclude all X-ray detections, as our primary focus is on constraining the LyC escape fraction from star-forming galaxies. In total, 11 galaxies from the non-detection sample and 2 from the LyC leaker candidate sample are detected in X-rays, one of the latter classified as a “Galaxy” and the other as an AGN, all of which are removed from further analysis. Thus our final sample comprises 42 LyC non-detection galaxies and 7 LyC leaker candidates.
The distribution of redshifts of this final sample is shown in Figure \ref{fig:z_distbn}. The redshift distribution of our sample also traces a previously reported overdensity at $z \sim 1.09$ in the Hubble Ultra Deep Field \citep{Boogard19}. 7 out of the 42 non-detections and 1 out of the 7 LyC leaker candidates lie within this narrow redshift range of $1.08 < z < 1.10$.
We display RGB images of the galaxies constructed using the HST F160W (red), F606W (green), and F435W (blue) bands of our sample in Figure \ref{fig:stamp_images1}. The newly identified LyC leaker candidate AUDFs\_F15450 in this work is highlighted in the bottom panel in Figure \ref{fig:stamp_images1}.

\section{Properties of the sample} \label{sec:properties}
We examine the distribution of several basic physical properties of our final sample, including the absolute UV magnitude at rest-frame $\sim1500~\text{\AA}$, stellar mass, and the observed UV continuum slope. 

For the stellar mass estimates, we adopt values from the ASTRODEEP-GS43 catalog \citep{Merlin21}, which provides SED-derived masses based on extensive multi-wavelength photometry, including HST/WFC3 imaging. These estimates are derived assuming a delayed exponential star formation history, the Bruzual and Charlot stellar population models \citep{Bruzual03}, a Salpeter IMF \citep{Salpeter55}, and include nebular emission lines. We adopt these values directly for our analysis.

To estimate the absolute UV magnitudes ($M_{\text{UV}}$), we perform aperture photometry in the HST/WFC3 F336W band using 1.4$''$ radius circular apertures. For galaxies in our redshift range, the F336W band provides a good proxy for the non-ionizing UV continuum around rest-frame 1500 $\text{\AA}$. We do not correct $M_{\text{UV}}$ for the internal dust extinction. 

We estimate the observed UV continuum slope, $\beta_{\text{obs}}$, by fitting a power-law function, $f_{\lambda} \propto \lambda^{\beta}$, to the flux densities measured in three HST bands: F336W, F435W, and F606W. These bands are chosen to approximately fall within the wavelength range defined by \citet{Calzetti94} for UV continuum slope measurements. For four galaxies lacking F336W coverage, we derive $\beta_{\text{obs}}$ using only the F435W and F606W bands. We also re-evaluate the $\beta_{\text{obs}}$ values for the LyC leaker candidates previously reported in \citet{Maulick25}, which were originally derived using SED model fluxes. Our updated measurements, based on observed photometry, yield systematically bluer slopes, with $\beta_{\text{obs}}$ values lower by approximately 0.5 compared to the earlier estimates.
\par
\begin{figure*}[ht!]
\nolinenumbers
\includegraphics[width=1\textwidth]{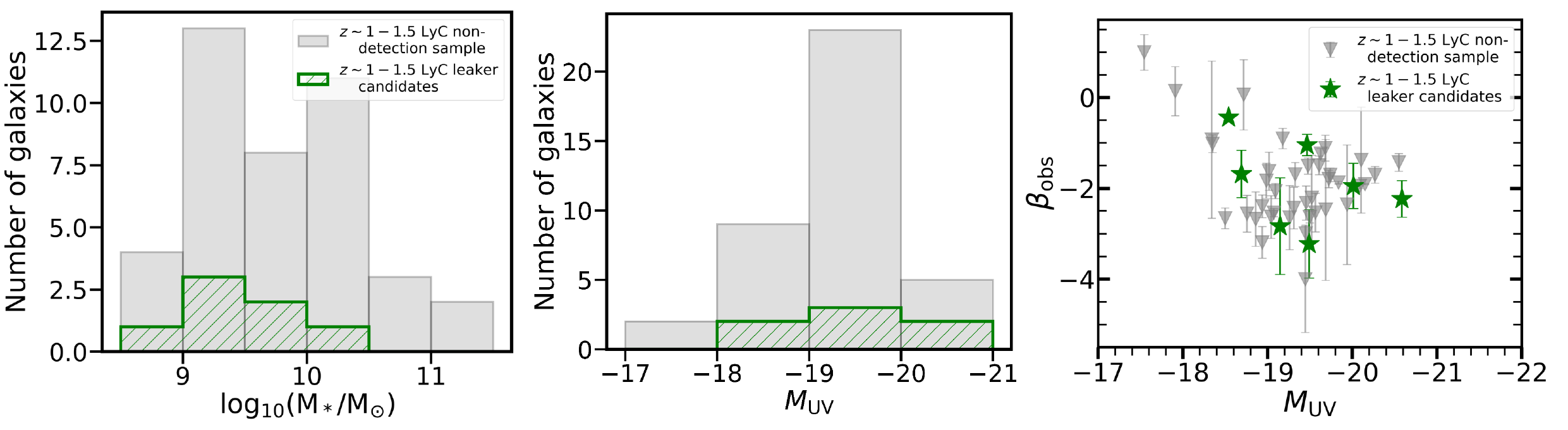} 
\caption{The first two panels (from left) show the distributions of stellar mass and absolute UV magnitude for our sample: grey filled histograms represent LyC non-detections, while green hatched histograms correspond to LyC leaker candidates. In the third panel, grey (non-detections) and green (leaker candidates) star markers show the relationship between the observed UV continuum slope ($\beta_{\text{obs}}$) and absolute UV magnitude for galaxies in our sample not detected in X-rays in the Chandra 7 Ms survey \citep{Luo17}. Colored circular markers indicate galaxies that are X-ray detected.} 
\label{fig:sample_prop}
\end{figure*}

 We display the distribution of the aforementioned properties in Figure \ref{fig:sample_prop}. Our sample spans a broad range of stellar masses. The $M_{\text{UV}}$ distribution indicates that a substantial fraction of our galaxies have luminosities comparable to $L_*$, consistent with the UV luminosity function at the redshifts probed in this study, as reported by \citet{Sharma22, Bhattacharya25}. For the LyC non-detections and LyC leaker candidates, the mean $\beta_{\text{obs}}$ values are $-1.84 \pm 0.95$ and $-1.91 \pm 0.97$, respectively. 

For all galaxies in our sample, we use H$\alpha$ fluxes and associated uncertainties from the 3D-HST \citep{Momvheva16} and CLEAR \citep{Simons23} catalogs. To compute quantities such as the star-formation rate surface density ($\Sigma_{\text{SFR}}$) derived from H$\alpha$ \citep{Kennicutt98}, we must correct for internal dust extinction. However, most galaxies in our sample lack significant H$\beta$ detections. We therefore estimate the stellar continuum color excess, $E(B-V)_{\text{stellar}}$, from the observed relation between the UV continuum slope and dust \citep{Reddy18}. We then convert this to the nebular color excess using the classical Calzetti relation, $E(B-V)_{\text{neb}} = 2.27 \times E(B-V)_{\text{stellar}}$ \citep{Calzetti94,Calzetti01}, while noting that this ratio exhibits substantial scatter \citep{Reddy20}. Finally, we apply the nebular curve of \cite{Reddy20} for the extinction correction.

\section{Image stacking analysis} 
\label{sec:stacking}

In this section, we describe our stacking analysis that we carry out in the far-ultraviolet F154W band image from the AUDFs Survey \citep{Saha24}. Unlike optical images, the F154W image is an instrinsically low-photon count image, meaning that the dominant source of noise, i.e., the sky background, exhibits a distinctly Poissonian distribution, as shown in \citet{Saha24,Pandey25}. Accurate estimation of the background and its associated noise is critical for the reliability of our stacking results. For the AUDFs observations, the UVIT CMOS detector's dark current and read noise are subdominant compared to the sky background \citep{Saha20}, and are therefore neglected in our stacking procedure. We perform uniform mean stacking of the sample in the F154W band. Specifically, we extract $20'' \times 20''$ sky-subtracted F154W band cutouts centered on the F606W centroids of each galaxy. To prevent contamination of the background by nearby bright sources in individual cutouts, we mask all objects outside a $1.4''$ radius aperture centered on each galaxy. Our target selection criteria minimize the likelihood of foreground contamination within 1.4$''$ of the cutout center.

For the neighboring objects detection and masking, we use the Python implementation of Source Extractor \citep{Bertin96}, \textit{sep} \citep{Barbary16}, with a $1\sigma$ pixel threshold and a minimum area of 5 connected pixels, noting that the F154W pixel scale is 1 pixel$\sim0.417''$. We convolve the image with a Gaussian filter of FWHM that matches that of the F154W PSF before source detection. This choice of filter, and of the detection parameters are motivated by extensive tests on pure Poisson background images \citep{Saha24, Maulick25}, as well as on the UVIT data. On average, our masking procedure results in the masking of $\sim5\%$ of pixels in each cutout. Masked pixels are then replenished by randomly sampling from neighboring unsegmented pixels, similar in spirit to the IRAF fixpix routine. For each cutout, we estimate and subtract any residual local background prior to stacking. The local background is estimated by placing 1.4$''$ radius apertures randomly across each individual masked cutout. We consistently find a positive residual background, suggesting a systematic offset in the background estimation of the F154W-band image. The stacking itself is then performed by computing the mean value of each pixel across the aligned cutouts. We measure the stacked flux within a $1.4''$ radius circular aperture placed at the center of the final stacked image. 

\begin{figure}[ht!]
\nolinenumbers
\includegraphics[width=0.48\textwidth]{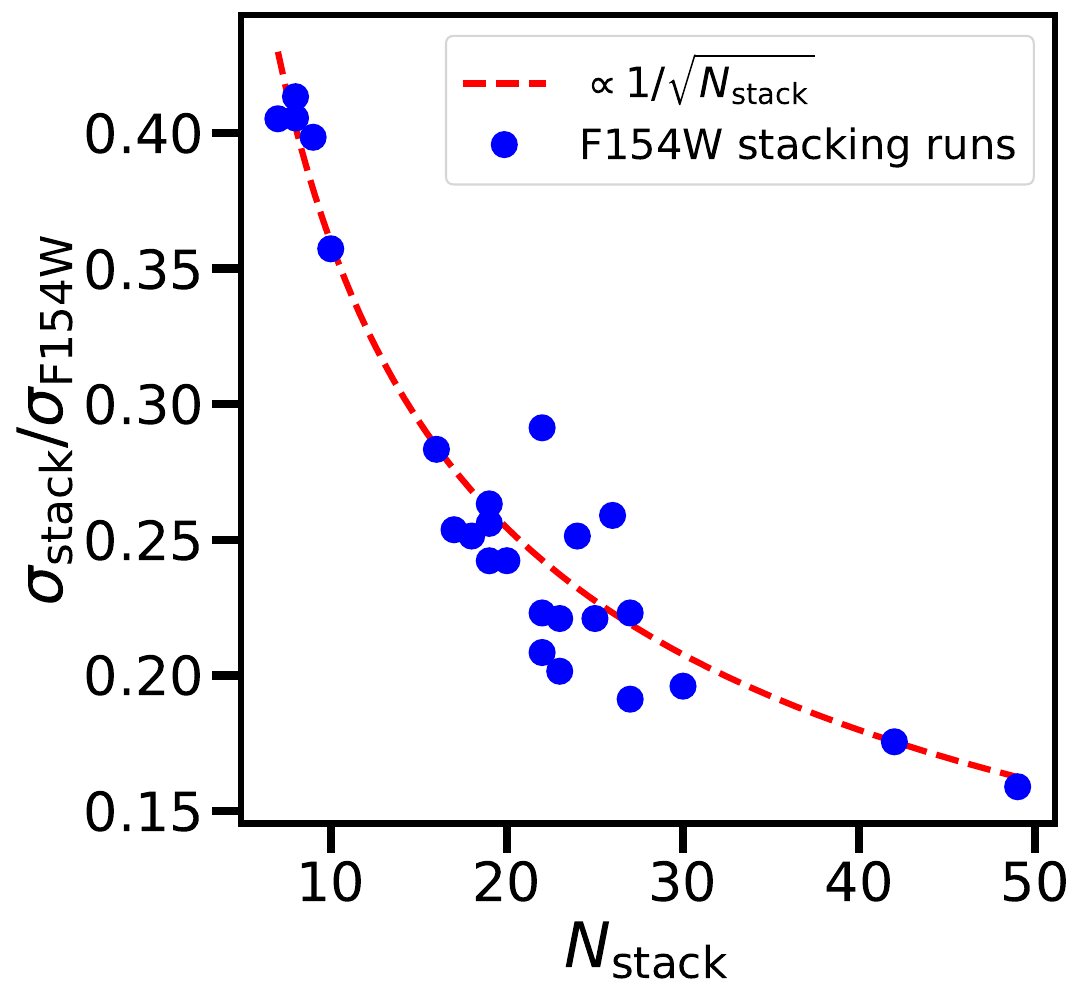} 
\caption{Noise in the stacked images as a function of the number of images combined ($N_{\text{stack}}$). The stacked noise, $\sigma_{\text{stack}}$, is normalized by the global average noise, $\sigma$, measured in the deepest region of the F154W image \citep{Saha24}, which effectively corresponds to the case of $N_{\text{stack}}=1$. Each blue point corresponds to a stacking run listed in Table \ref{tab:stack_results}, while the red dashed curve shows the expected $1/\sqrt{N_{\text{stack}}}$ scaling.} 
\label{fig:stack_sig_scale}
\end{figure}

To assess the significance of the stacked signal, we adopt a random aperture sampling approach. We place non-overlapping apertures of the same size (1.4$''$ radius circular apertures) used to measure the stacked signal at the center, at random locations on the stacked image (excluding the center) to estimate the noise, following practices in \citet{Zheng2006, Rafelski15, Prichard22, Wang25}. This also allows us to quantify the depth of the stacked image in terms of the flux distribution from these empty apertures.
Finally, we repeat the entire stacking procedure multiple times to account for the stochasticity introduced by both the pixel replenishment process and the random placement of background apertures, ensuring robust error estimation.
In Figure \ref{fig:stack_sig_scale}, we present how the estimated noise in our stacked images varies with the number of images combined ($N_{\text{stack}}$). The measurements (blue points), each corresponding to a stacking run listed in Table \ref{tab:stack_results}, are broadly consistent with the theoretical expectation for mean stacking, where the noise decreases as $1/\sqrt{N_{\text{stack}}}$ (red dashed curve).

\par
\begin{figure*}[ht!]
\nolinenumbers
\includegraphics[width=1\textwidth]{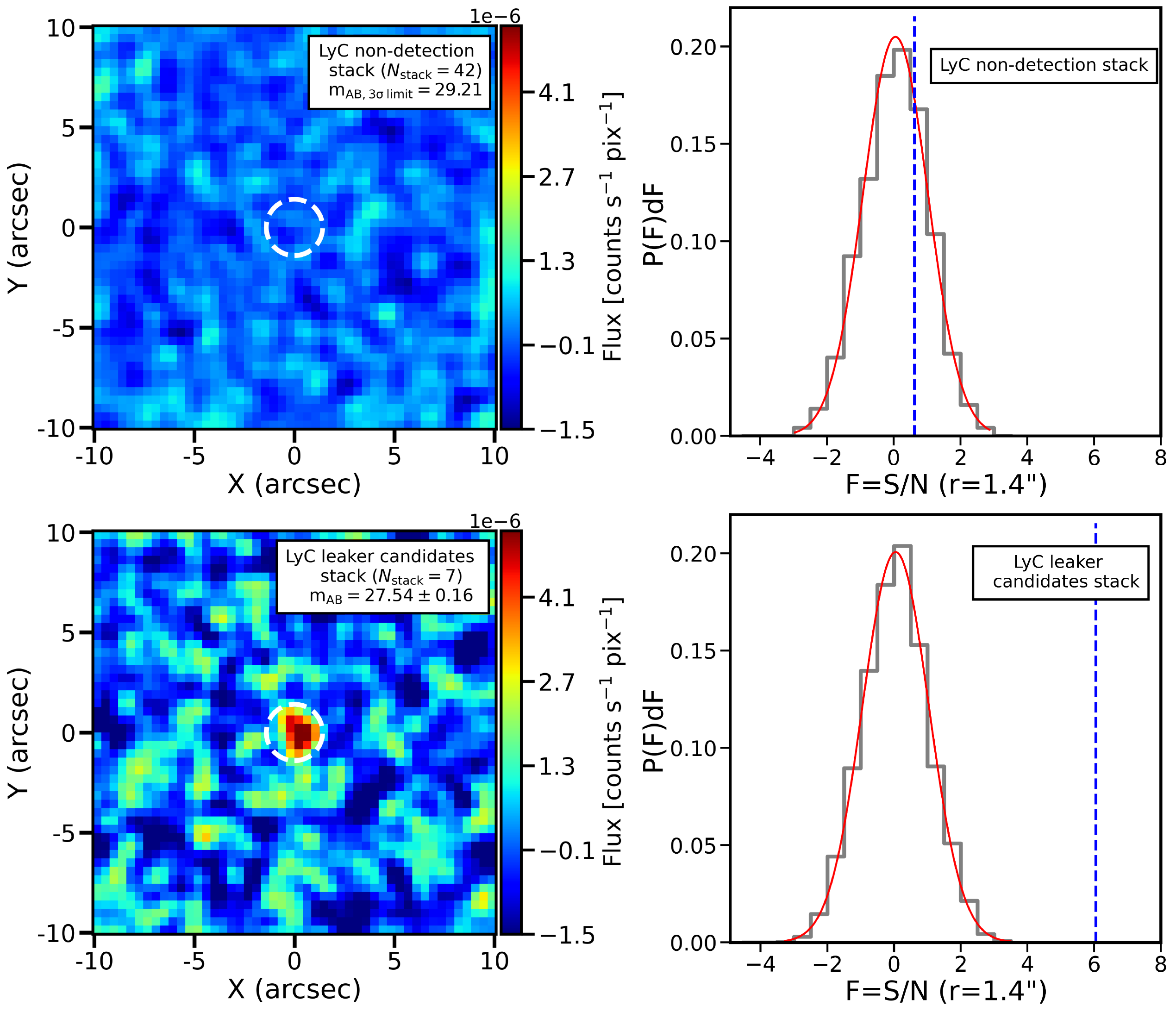} 
\caption{Top panel: The F154W-band stacked image created from 42 individual $20'' \times 20''$ cutouts centered on galaxies in the LyC non-detection sample. The images have been smoothed using a Gaussian kernel having a standard deviation of nearly 1 pixel. $N_{\text{stack}}$ denotes the number of galaxies contributing to the stacked image.
Right panel: Normalized distribution of the normalized fluxes measured within randomly placed apertures across the stacked image, excluding the central region. The fluxes are normalized by the standard deviation of a Gaussian fitted to the distribution of fluxes, denoted here by the solid red curve. The dashed blue line indicates the significance of the flux measured within the central white dashed aperture.
Bottom panel: Same as above, but for the stack of F154W cutouts corresponding to the 7 LyC leaker candidates.} 
\label{fig:stack_image}
\end{figure*}

 We present in Figure~\ref{fig:stack_image} the stacked F154W images (left panels) for the 42 LyC non-detections (top row) and the 7 LyC leaker candidates (bottom row), along with the corresponding significance distributions (right panels), evaluated with respect to fluxes measured in randomly placed empty apertures across the stacked images. The stack of the non-detections does not yield a significant detection, with a $3\sigma$ limiting depth of 29.21 mag within a 1.4$''$ aperture. In contrast, the stack of the LyC leaker candidates shows a $\sim6\sigma$ detection relative to the background. When stacking all 49 galaxies in our sample, we obtain a marginal detection with a significance of 2.40$\sigma$, indicating that the stacked signal is primarily driven by the leaker candidates.

To assess the robustness of our procedure, particularly the object detection, masking, and residual background subtraction in individual cutouts, we use deep JWST/NIRCam and HST/WFC3 images to select 49 “empty-sky” positions, offset from our targets, matching the size of our entire sample. By ''empty-sky'' we mean locations that contain no detectable object in the JWST and HST optical/infrared imaging. We then perform the same stacking analysis, treating these positions as the central targets. This yields a significance of –0.58 at the center relative to the stacked background, indicating that our stacking procedure does not produce spurious positive signals inconsistent with expectations for pure background fluctuations.

\section{Results} 
\label{sec:results}

We employ our stacking analysis to investigate the escape of ionizing photons from various subsamples of our parent galaxy sample, defined based on physical properties believed to influence LyC escape. Specifically, we aim to quantify the observed ratio of LyC to non-ionizing UV flux density, $({F}_{\lambda,\text{LyC}}/{F}_{\lambda,\text{UV}})_{\text{obs}}$, using the F154W and HST F336W bands as proxies, respectively. This ratio offers a model-independent measure and facilitates direct comparison with previous studies in the literature. For each bin, we also estimate the average absolute UV magnitude, $\langle M_{\text{UV}}\rangle$ for that bin, by averaging the luminosity density inferred from the F336W band flux. We emphasize that this estimate is not corrected for internal dust extinction. 
It is worth noting that the F154W band typically probes bluer LyC wavelengths (around $715\: \text{\AA}$) compared to most LyC studies, which focus on wavelengths near the Lyman limit. However, importantly, prior studies targeting similar redshift ranges such as those using the HST ACS/SBC F150LP filter \citep{Siana07,Siana10,Alavi20} and the GALEX FUV band \citep{Rutkowski16} also sample rest-frame LyC wavelengths comparable to those probed by the F154W band. We correct our upper limits and obtained values of $({F}_{\lambda,\text{LyC}}/{F}_{\lambda,\text{UV}})_{\text{obs}}$ for the average IGM attenuation at $z\sim1.2$ using the models of \citet{Inoue14} that implies a mean transmission of 0.63 at this redshift \citep[see also][]{Dhiwar24}. We present our constraints corresponding to the various bins described in the following subsections, in Table \ref{tab:stack_results}.

\subsection{Stellar mass}
One of the key questions relevant to the Epoch of Reionization (EoR) is whether the dominant contributors to the ionizing photon budget were low-mass or high-mass galaxies. While the debate remains open \citep{Naidu20}, the prevailing view is that low-mass galaxies play a dominant role due to their sheer abundance \citep{Finkelstein19,Atek24}. In addition to their numbers, low-mass galaxies are also thought to possess interstellar media (ISM) more favorable to LyC escape, potentially resulting in higher escape fractions \citep{Ma20,Rosdahl22}. We investigate this possibility in Section \ref{sec:discussion}.

High-resolution simulations suggest that the LyC escape fraction peaks at stellar masses of $\sim10^8 \:\text{M}_{\odot}$ \citep{Ma20}, or even earlier \citep{Rosdahl22}, and declines thereafter. Given observational limitations, we divide our sample into two bins, low mass and high-mass, using $10^{9.5} \:\text{M}_{\odot}$ as the threshold. This choice balances the sample sizes and ensures comparable stacking depths across the two bins. When stacking only the non-detections, we find no significant signal in either the low- or high-mass bins. However, upon including our 7 LyC leaker candidates, four in the low-mass bin and three in the high-mass bin, we detect a signal in the low-mass bin with a S/N=4.35 relative to the stacked background. The high-mass bin remains undetected.

\subsection{Observed UV continuum slope}
We next consider the observed UV continuum slope, $\beta_{\text{obs}}$, to define subsamples. This choice is motivated by studies at lower redshifts, the LzLCS in particular \citet{Flury22a}, that find an anti-correlation between $\beta_{\text{obs}}$ and the LyC escape fraction \citep{Chisholm22}. Blue UV continuum slopes are typically associated with young stellar populations and low dust attenuation, conditions favorable for both the production and escape of LyC photons \citep{Zackrisson13}. Furthermore, the average UV continuum slope of star-forming galaxies has been observed to become progressively bluer with increasing redshift \citep{Bouwens14, Mondal23}, with recent JWST studies \citep{Cullen23} showing that, by $z \sim 11.5$, it approaches the intrinsic blue limit expected for dust-free stellar populations. Similar to the stellar mass binning, we divide our sample into two bins based on $\beta_{\text{obs}}$, adopting a cutoff at $\beta_{\text{obs}} = -1.8$ to separate the blue and red UV continuum slope subsamples. We do not find any significant detection in the stacks compirsing of just the non-detections in either bin. However, upon including the LyC leaker candidates, we obtain a marginal detection with a significance of 2.89 in the blue bin ($\beta_{\text{obs}} < -1.8$).

\subsection{Orientation}
In Maulick et al. (2025b, submitted), we present tentative evidence of the detection of LyC being more likely from face-on galaxies, than edge-on galaxies. However, we did suspect that the results of the experiment may have been biased by the presence of AGN in that sample. 

Here, we create two ''near edge-on'' and ''near face-on'' bins in the same manner as in Maulick et al. (2025b, submitted). Namely, we utilize the structural fits to the galaxies from the catalog of \cite{vanderWel12} and select for disk galaxies using a Sérsic index $n<2$ as the selection for disk galaxies. We then estimate their angle of inclination using their observed axis ratios using the classical Hubble formula \citep{Hubble26}: $\text{cos}^2(i)=(q^2-q_0^2)/(1-q_0^2)$, where where $q=b/a$ is the observed axis ratio and for simplicity, we adopt $q_0=0.2$ as the intrinsic thickness of an edge-on disk \citep{Padilla08}. Galaxies with an inclination angle $i<45^{\circ}$ are classified as ''near face-on'' and galaxies with $i>45^{\circ}$ are classified as ''near edge-on''. We then perform our stacking analysis on these two bins. Contrary to our expectations, we do not detect any significant signal in the ''near face-on'' bin. 

\subsection{Compactness and star-formation rate surface density} \label{sec:compact}
At low redshifts ($z \sim 0.2$–0.4), compact galaxies with high star-formation rate surface densities have been found to be more likely to exhibit significant LyC leakage \citep{Flury22b, MarquesChaves22,Jaskot24b, Carr25}. This trend is supported by both theoretical and semi-analytic arguments, which suggest that both radiative and mechanical feedback is more effective at clearing gas in such systems \citep{Cen20, Carr25}. Empirical studies further show that galaxies hosting high-velocity galactic winds, which could facilitate LyC escape, are preferentially associated with high star-formation rate surface densities ($\Sigma_{\text{SFR}} > 0.1\ \text{M}_{\odot} \:\text{yr}^{-1}\ \text{kpc}^{-2}$, \cite{Heckman02}). This motivates us to construct subsamples based on galaxy compactness and star-formation rate surface density.

To quantify the compactness, we estimate the half-light radii ($r_{50}$) of all galaxies in our sample using a curve-of-growth analysis in the HST ACS/WFC F435W band. Compactness is not a well-defined quantity, as it depends on both instrument resolution and galaxy redshift. Since our sample spans only a narrow range in angular scale ($\sim8$–$8.5\ \text{kpc}/''$), we define 'compact' galaxies as those with $r_{50} < 2\times$ the PSF FWHM in the F435W band (PSF FWHM $\sim$ 2 pixels). For our dataset, this criterion corresponds exactly to a physical threshold of $r_{50} < 2\ \text{kpc}$, and galaxies with larger radii are classified as “extended.”  Similar to the low-mass and $\beta_{\text{obs}}$ stacks, we obtain a marginal detection ($\text{S/N}_{\text{stack}} \sim 2.45$) in the “compact” bin, but only when the LyC leaker candidates that meet the compactness criterion are included. No significant signal is detected in the “extended” bin.

\begin{figure*}[ht!]
\nolinenumbers
\centering
\textbf{Compact-high $\Sigma_{\text{SFR}}$-blue UV continuum slope LyC non-detection sample}\\[1pt]
\vspace{-0.1cm}
\includegraphics[width=1\textwidth]{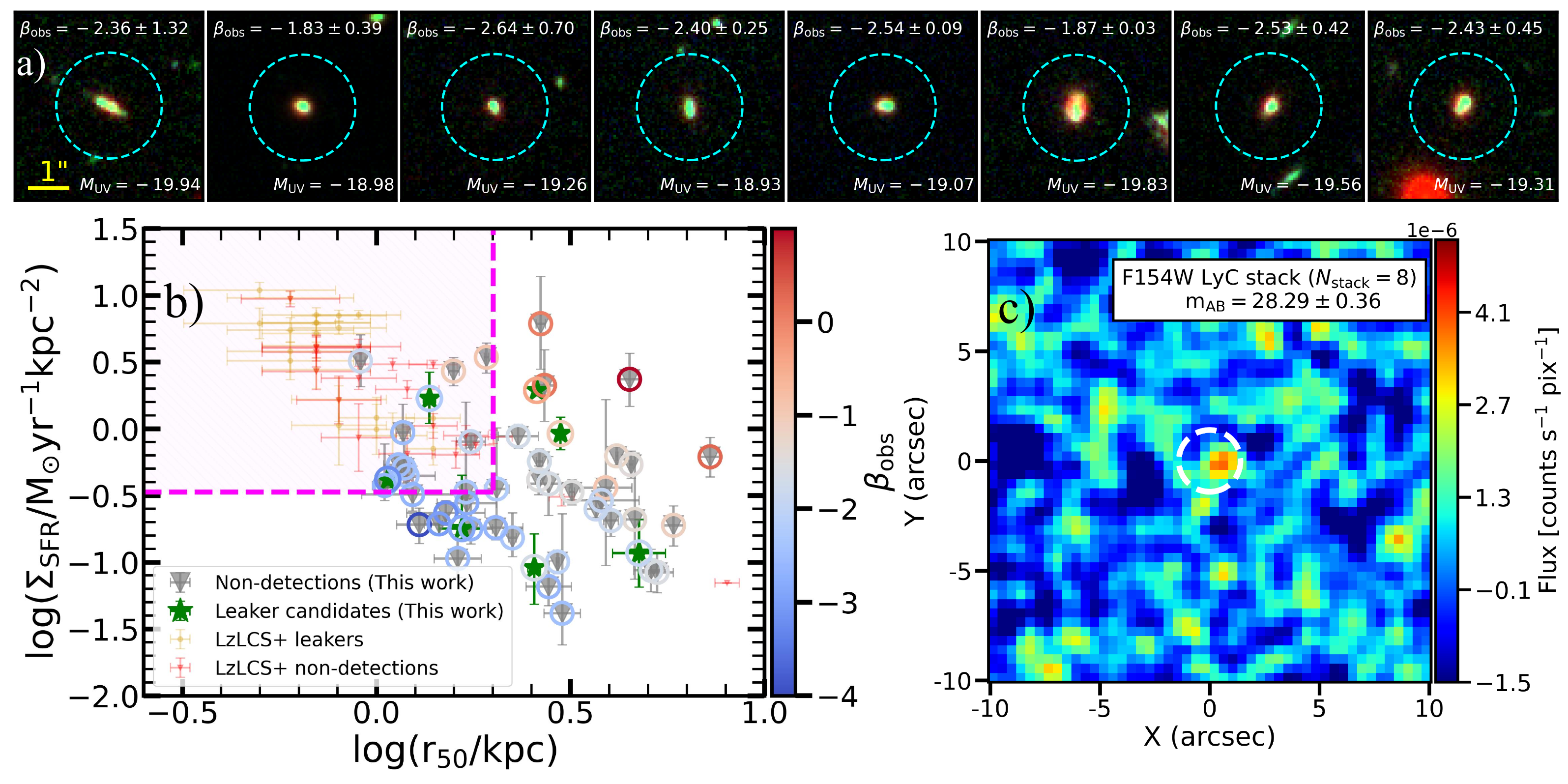} 
\caption{Panel a shows the RGB images (same as in Figure \ref{fig:stamp_images1}) of the 8 galaxies in our LyC non-detection sample that satisfy the compact, high star-formation rate surface density, and blue UV continuum slope selection criteria defined in Section \ref{sec:compact}. Their $\beta_{\text{obs}}$ and $M_{\text{UV}}$ values are provided in white text. Panel b displays our full sample in the $r_{50}$–$\Sigma_{\text{SFR}}$ plane, along with LyC detected galaxies (golden circles) and non-detections (red inverted triangles) from the LzLCS+ sample for comparison. The horizontal and vertical dashed magenta lines mark our cuts in compactness and high $\Sigma_{\text{SFR}}$, respectively. The hatched magenta-shaded region highlights the parameter space defined by these cuts. Colored rings around the galaxies in our sample indicate their $\beta_{\text{obs}}$ values. Panel c shows the stacked F154W image ($20'' \times 20''$ field) of the 8 galaxies in Panel a. The flux within the central white dashed aperture of radius 1.4$''$ corresponds to S/N$\sim$3.} 
\label{fig:high_sigsfr_compact}
\end{figure*}

We next construct subsamples based on the 
the star-formation rate surface density ($\Sigma_{\text{SFR}}$).
We estimate $\Sigma_{\text{SFR}}$ for our sample using, the formula, $\Sigma_{\text{SFR}}=\text{SFR}_{\text{H}\alpha}/(2 \pi\text{r}_{50}^{2})$, where for obtaining $\text{SFR}_{\text{H}\alpha}$, we use the dust-corrected H$\alpha$ luminosity \citep{Kennicutt98}. We display our sample in the $r_{50}-\Sigma_{\text{SFR}}$ plane in Figure \ref{fig:high_sigsfr_compact} (Panel b). We also compare our sample with galaxies of the LzLCS+. We plot the LzLCS+ galaxies in the same plane using half-light radii measurements from \cite{LeReste25}. On average, the galaxies in our sample are more extended than the LzLCS+ sample. We note that the LzLCS+ sample selection is biased towards the Green Pea \citep{Cardamone09} kind of galaxies given the understanding that these systems better facilitate ionizing photon escape \citep{Izotov2016}.
Given the uncertainties in our $\Sigma_{\text{SFR}}$ calculations due to the dust, we adopt the median $\Sigma_{\text{SFR}}\sim0.34\:\text{M}_{\odot}\:\text{yr}^{-1}\:\text{kpc}^{-2}$ of our sample as a cut to distinguish between high and low $\Sigma_{\text{SFR}}$ galaxies in our sample. The results of the $\Sigma_{\text{SFR}}$ stacks are similar to those of the compact/extended stacks with a marginal detection being detected only in the high $\Sigma_{\text{SFR}}$ bin that includes the LyC leaker candidates.

In the final bin, we combine multiple parameters to isolate galaxies that simultaneously satisfy all three criteria: being compact, having high $\Sigma_{\text{SFR}}$, and blue UV continuum slopes (last row in Table \ref{tab:stack_results}). The compact–high $\Sigma_{\text{SFR}}$ region is marked by the hatched magenta shading in Panel b of Figure \ref{fig:high_sigsfr_compact}. We note that the majority of the LzLCs+ sample falls within this region. Adding the UV continuum slope cut further selects objects expected to host young stellar populations and/or low dust content. Applying these combined criteria yields 8 galaxies from our LyC non-detection sample, highlighted in Panel a of Figure \ref{fig:high_sigsfr_compact}.

Stacking these 8 galaxies in the F154W band (Panel c of Figure \ref{fig:high_sigsfr_compact}) produces a $\sim3\sigma$ detection. Notably, across all our property-based bins, this is the only subsample composed exclusively of LyC non-detections that yields a significant signal. The mean S/N of these eight galaxies from forced aperture photometry is $\sim1.05$, and thus the S/N of the stack is consistent with the expected  scaling by a factor of $\sqrt{8}$. One galaxy exhibits a relatively high S/N of $\sim3.4$. Although such a source might be expected to fall within our LyC-leaker sample, on inspection, we find that despite its high photometric S/N, it is not detected by Source Extractor in the F154W band. According to our LyC leaker candidate definition, this object is therefore classified as a non-detection. Excluding this source and stacking the remaining seven non-detections yields a stack S/N of $\sim2$, indicating that this object has a noticeable but not dominant effect on the stacked signal. Finally, including the two LyC leaker candidates that also meet these criteria increases the detection significance to $\sim4.4$.

\begin{table*}[ht]
    
    \caption{Our main stacking results for different galaxy subsamples. Columns from left to right: criterion for the sample, number of galaxies in the bin, stacked F154W magnitude or its 3$\sigma$ lower limit (no aperture correction), mean absolute UV magnitude, the observed LyC-to-nonionizing UV flux density ratio or its 3$\sigma$ upper limit (LyC flux is aperture- and IGM-corrected), and the S/N at the stack center, indicated in boldface when greater than 3. Fluxes and the S/N are measured in circular apertures of radius 1.4$''$. The S/N is estimated using the method outlined in Section \ref{sec:stacking}. Each row, corresponding to a galaxy property under examination, is divided into two sub-bins: one containing only LyC non-detection galaxies (ND), and the other containing both LyC non-detection galaxies and LyC leaker candidates (ND+L) that satisfy the given criterion.}
\label{tab:stack_results}
    \scalebox{0.75}
    {\hspace{-2cm} \renewcommand{\arraystretch}{1.8} \setlength{\tabcolsep}
    {7pt}\begin{tabular}{|lcccccccccc|}
    \hline
        {Sample} & \multicolumn{2}{c}{$N_{\text{stack}}$} & \multicolumn{2}{c}{$\text{m}_{\text{AB,F154W stack}}$} & \multicolumn{2}{c}{$\langle M_{\text{UV}}\rangle$} & \multicolumn{2}{c}{$({F}_{\lambda,\text{LyC}}/{F}_{\lambda,\text{UV}})_{\text{obs}}$} & \multicolumn{2}{c|}{(S/N)$_{\text{stack}}$} \\
        \hline
        
        \hline
        All nondetections (ND) & & \hspace{-1cm}42 &&\hspace{-2cm}$>29.21$ & &\hspace{-1.5cm}-19.45 & &\hspace{-2cm}$<0.12$ & &\hspace{-1.5cm} 0.62 \\
        All LyC leaker candidates (L)   & & \hspace{-1cm}7 &&\hspace{-2cm}$27.54\pm0.16$ & &\hspace{-1.5cm}-19.62 & &\hspace{-2cm}$0.50\pm0.08$ & &\hspace{-1.5cm} \textbf{6.05} \\
        All galaxies (ND+L) & & \hspace{-1cm}49 &&\hspace{-2cm}$29.56\pm0.45$ & &\hspace{-1.5cm}-19.48 & &\hspace{-2cm}$0.09\pm0.04$ & &\hspace{-1.5cm} 2.40   \\
        \hline & ND & ND+L & ND & ND+L & ND & ND+L & \hspace{0.3cm}ND & ND+L & ND & ND+L \\
        Low-mass ($\text{log}_{10}(\text{M}_{*}/\text{M}_{\odot})<9.5$) & 18 & 22 & $>28.82$ & $28.62\pm0.25$ & $-19.23$ & $-19.34$ & $<0.22$ & $0.24\pm0.06$ & $1.54$ & \textbf{4.35} \\
        High-mass ($\text{log}_{10}(\text{M}_{*}/\text{M}_{\odot})>9.5$) & 24 & 27 & $>28.82$ & $>28.95$ & $-19.60$ & $-19.59$ & $<0.15$ & $<0.14$ & $-0.49$ & 0.33  \\
        \hline
        $\beta_{\text{obs}}<-1.8$ & 23 & 27 & $>29.06$ & $28.91\pm0.30$ & $-19.42$ & $-19.52$ & $<0.15$ & $0.16\pm0.05$ & $1.67$ & \textbf{3.63} \\
        $\beta_{\text{obs}}>-1.8$ & 19 & 22 & $>28.80$ & $>28.95$ & $-19.50$ & $-19.44$ & $<0.16$ & $<0.21$ & $-0.75$ & 0.35 \\
        \hline
        Near face-on & 8 & 9 & $>28.28$ & $>28.32$ & $-19.64$ & $-19.56$ & $<0.25$ & $<0.26$ & $-1.58$ & -0.76 \\
        Near edge-on & 16 & 19 & $>28.69$ & $>28.77$ & $-19.20$ & $-19.19$ & $<0.26$ & $<0.23$ & $0.16$ & 1.70 \\
        \hline
        Compact ($r_{50}\leq2$ kpc) & 17 & 19 & $>28.81$ & $29.08\pm0.44$ & $-19.23$ & $-19.24$ & $<0.22$ & $0.17\pm0.07$ & $1.09$ & 2.45 \\
        Extended ($r_{50}>2$ kpc) & 25 & 30 & $>28.96$ & $>29.09$ & $-19.59$ & $-19.62$ & $<0.13$ & $<0.12$ & $-0.09$ & 1.89 \\
        \hline
        High $\Sigma_{\text{SFR}}\:(>0.34\:\text{M}_{\odot}\:\text{yr}^{-1}\:\text{kpc}^{-2})$ & 21 & 25 & $>28.66$ & $29.10\pm0.48$ & $-19.42$ & $-19.48$ & $<0.21$ & $0.14\pm0.07$ & $0.96$ & 2.25 \\
        Low $\Sigma_{\text{SFR}}\:(<0.34\:\text{M}_{\odot}\:\text{yr}^{-1}\:\text{kpc}^{-2})$ & 21 & 24 & $>28.86$ & $>28.96$ & $-19.50$ & $-19.49$ & $<0.17$ & $<0.15$ & $-0.08$ & 1.30 \\
        \hline
        \hline
        
        $r_{50}\leq2$ kpc, $\Sigma_{\text{SFR}}\:>0.34\:\text{M}_{\odot}\:\text{yr}^{-1}\:\text{kpc}^{-2}$, and $\beta_{\text{obs}}<-1.8$ & 8 & 10 & $28.29\pm0.36$ & $28.02\pm0.24$ & $-19.42$ & $-19.62$ & $0.30\pm0.10$ & $0.35\pm0.08$ & \textbf{3.03} & \textbf{4.41} \\ \hline
    \end{tabular}}
    
\end{table*}

\section{Discussion} 
\label{sec:discussion}
The majority of galaxies in our sample are not detected in the F154W band imaging. Among the seven LyC leaker candidates in our sample, we find a striking diversity in their physical properties. For instance, AUDFs\_F15450 (bottom panel of Figure \ref{fig:stamp_images1}) has a stellar mass of $\log_{10}(\text{M}_{*}/\text{M}_{\odot})\sim9.29$ and exhibits an exceptionally blue UV continuum slope ($\beta_{\text{obs}}\sim-3.2$), consistent with extremely young, dust-free stellar populations \citep{Katz25}. Owing to the high S/N measurements of the $[\text{O III}] \lambda\lambda4959,5007$ and $[\text{O II}] \lambda\lambda3727,3729$ lines from the 3D-HST and MUSE-WIDE catalogs respectively, we estimate an O32 ratio of 6.5 for this system, suggestive of LyC escape facilitated by a density-bounded ISM \citep{Nakajima14}. This galaxy also lies within our bin of compact, high-$\Sigma_{\text{SFR}}$ objects with blue UV continuum slopes (Section \ref{sec:compact}). In contrast, the spiral galaxy reported in Maulick et al. (2025b, submitted) is a comparatively massive system ($\log_{10}(\text{M}_{*}/\text{M}_{\odot})\sim10.32$) with a red UV continuum slope ($\beta_{\text{obs}}\sim-0.4$), and shows no evidence of an elevated ionization state (Maulick et al. (2025b, submitted)). These contrasting cases suggest that the mechanisms enabling LyC escape may differ substantially across individual galaxies.

Beyond these individual cases, our stacking analysis reveals population-level trends. In particular, the LyC detection in the low stellar mass bin, combined with its average UV luminosity (uncorrected for dust), leads to a higher observed LyC-to-nonionizing UV flux density ratio than in the high stellar mass bin, for which we only establish upper limits. However, this comparison does not directly translate into a physically meaningful measure of the efficiency of ionizing photon escape. To place this result in context, we must first translate the observed flux density ratio into the more physically relevant quantity, i.e., the LyC escape fraction.
The starting point is the definition of the relative escape fraction \citep{Steidel01},

\begin{equation} \label{eq:fesc_rel}
f_{\text{esc,rel}}=\frac{({F}_{\lambda,\rm{LyC}}/{F}_{\lambda,\rm{UV}})_{\rm{obs}}}{({L}_{\lambda,{\rm{LyC}}}/{L}_{\lambda,\rm{UV}})_{\rm{int}}},
\end{equation}
where ${({L}_{\lambda,\rm{LyC}}/{L}_{\lambda,\rm{UV}})_{\rm{int}}}$ denotes the intrinsic LyC-to-nonionizing UV luminosity density ratio predicted by stellar population models. Since dust attenuation is the only sink for non-ionizing UV photons, the relative escape fraction can be converted into the absolute escape fraction as,
\begin{equation} \label{eq:fesc_rel_to_abs}
f_{\text{esc,abs}}=f_{\text{esc,rel}}\times10^{-0.4A_{\lambda}(\text{UV})},
\end{equation}
where $A_{\lambda}=k_{\lambda}E(B-V)_{\text{stellar}}$ corresponds to the dust extinction at the UV wavelength.

We find that the high-mass bin is significantly dustier than the low-mass bin, with $\langle E(B-V)_{\text{stellar}}\rangle^{\text{high-mass}}\sim0.29$ compared to $\langle E(B-V)_{\text{stellar}}\rangle^{\text{low-mass}}\sim0.04$. Here, we emphasize again that $E(B-V)_{\text{stellar}}$ is derived from the empirical relation between the UV continuum slope and $E(B-V)_{\text{stellar}}$ \citep{Reddy18}. If the intrinsic LyC-to-nonionizing UV ratio, ${({L}_{\lambda,\rm{LyC}}/{L}_{\lambda,\rm{UV}})_{\rm{int}}}$, is not dramatically different between the two mass bins, then this dust difference and the observed flux ratios imply that $\langle f_{\text{esc,abs}}\rangle^{\text{low-mass}}>\langle f_{\text{esc,abs}}\rangle^{\text{high-mass}}$.

However, we caution that ${({L}_{\lambda,\rm{LyC}}/{L}_{\lambda,\rm{UV}})_{\rm{int}}}$ is sensitive to stellar age, metallicity, initial mass function, and star-formation history \citep{Siana07,Chisholm19}, which may differ systematically between low- and high-mass galaxies. These in turn would influence the estimate of the absolute escape fraction.
For instance, the intrinsic value ${({L}_{\lambda,\rm{LyC}}/{L}_{\lambda,\rm{UV}})_{\rm{int}}}$ predicted by the BPASS \citep{Stanway18} binary models with a Salpeter IMF and a young (5 Myr) single age burst population, is 1.89 for the most metal-poor case ($Z=10^{-5}Z_{\odot}$). Assuming the Calzetti law for dust extinction \citep{Calzetti01}, this translates to $\langle f_{\text{esc,abs}}\rangle^{\text{low-mass}}\sim0.09$ for our measured $({F}_{\lambda,\rm{LyC}}/{F}_{\lambda,\rm{UV}})_{\text{obs}}$. On the other hand, for the same stellar population but with solar metallicity, ${({L}_{\lambda,\rm{LyC}}/{L}_{\lambda,\rm{UV}})_{\rm{int}}}=0.72$, which translates to $\langle f_{\text{esc,abs}}\rangle^{\text{low-mass}}\sim0.23$. From the stack of all galaxies, and adopting a mean stellar reddening of $\langle E(B-V)_{\text{stellar}}\rangle \sim 0.18$ for the sample, we derive extremely low absolute escape fractions of $\langle f_{\text{esc,abs}}\rangle\sim0.01$ (metal-poor) and $\sim0.02$ (solar metallicity), assuming the same intrinsic ratios as above for a 5 Myr single-age burst binary population. Comparable values of $\langle f_{\text{esc,abs}}\rangle$ are obtained when adopting the commonly used intrinsic ratio ${({L}_{\lambda,\rm{LyC}}/{L}_{\lambda,\rm{UV}})_{\rm{int}}}=0.92$, corresponding to LyC wavelengths near $\sim900 \:\text{\AA}$ (i.e., roughly one-third in frequency space; \citealt{Steidel01,Siana07,Grazian16,Kerutt23}), as this value lies intermediate to those adopted from BPASS in our analysis. {From the stack of the 42 non-detections, that yields no LyC signal, we infer an upper limit of $\langle f_{\text{esc,abs}}\rangle < 0.03$ when considering the same range of stellar population models.} These estimates, however, carry significant additional uncertainty, stemming from both the dust correction and the measurement errors on the stacked signal. A detailed assessment of these effects, including constraints on the stellar populations of our sample to better determine the intrinsic ratio ${({L}_{\lambda,\rm{LyC}}/{L}_{\lambda,\rm{UV}})_{\rm{int}}}$, is deferred to future work.

Our results suggest that galaxies that are compact, exhibit high $\Sigma_{\text{SFR}}$, and have bluer UV continuum slopes preferentially leak ionizing photons. Considering each of these properties in isolation yields a detection only when including the LyC leaker candidates that satisfy it. However, when all three properties are combined, we obtain a detection, albeit marginally significant, even when restricting the sample to objects that are individually undetected in LyC. This suggests that these galaxies may share underlying physical conditions that facilitate LyC escape. LyC leakage can occur either anisotropically, through channels carved by supernovae and stellar winds, or isotropically, in density-bounded systems \citep{Zackrisson13}. More recently, two-stage burst models that combine elements of both mechanisms have also been proposed \citep{Flury25}.
It is therefore plausible that some of our non-detections, including high-mass extended galaxies, still host sites of LyC leakage, but the escape of LyC photons is along unfavorable lines of sight. In such cases, a random distribution of a few anisotropic escape channels would not yield detections, even in stacked analyses. Furthermore, for extended galaxies, our stacking method that is anchored to the optical centroids may misalign the true sites of LyC production and escape (e.g., off-centred clumps within different galaxies). Consequently, LyC emission from individual clumps could be diluted in the stack, contributing to the non-detections.
By contrast, the LyC escape observed in compact, high-$\Sigma_{\text{SFR}}$ galaxies with blue-UV continuum slopes may instead reflect a more isotropic, density-bounded scenario.
Although galaxies that satisfy all three criteria constitute only $\sim 20\%$ of our sample, they appear more common at higher redshifts, closer to the EoR \citep{Mascia25}. Using survival analysis on the Low-redshift Lyman Continuum Survey (LzLCS), \citet{Jaskot24b} identified compactness, $\beta_{\text{obs}}$, and $\Sigma_{\text{SFR}}$ as among the strongest predictors of LyC escape, with $\beta_{\text{obs}}$ and $\Sigma_{\text{SFR}}$ ranking among their top three predictors of the escape fraction. Our findings are therefore consistent with these results, reinforcing the picture that high-redshift analogs of such galaxies are likely to exhibit stronger LyC leakage.

\subsection{Comparison with other LyC studies}
\begin{figure*}[ht!]
\nolinenumbers
\includegraphics[width=1\textwidth]{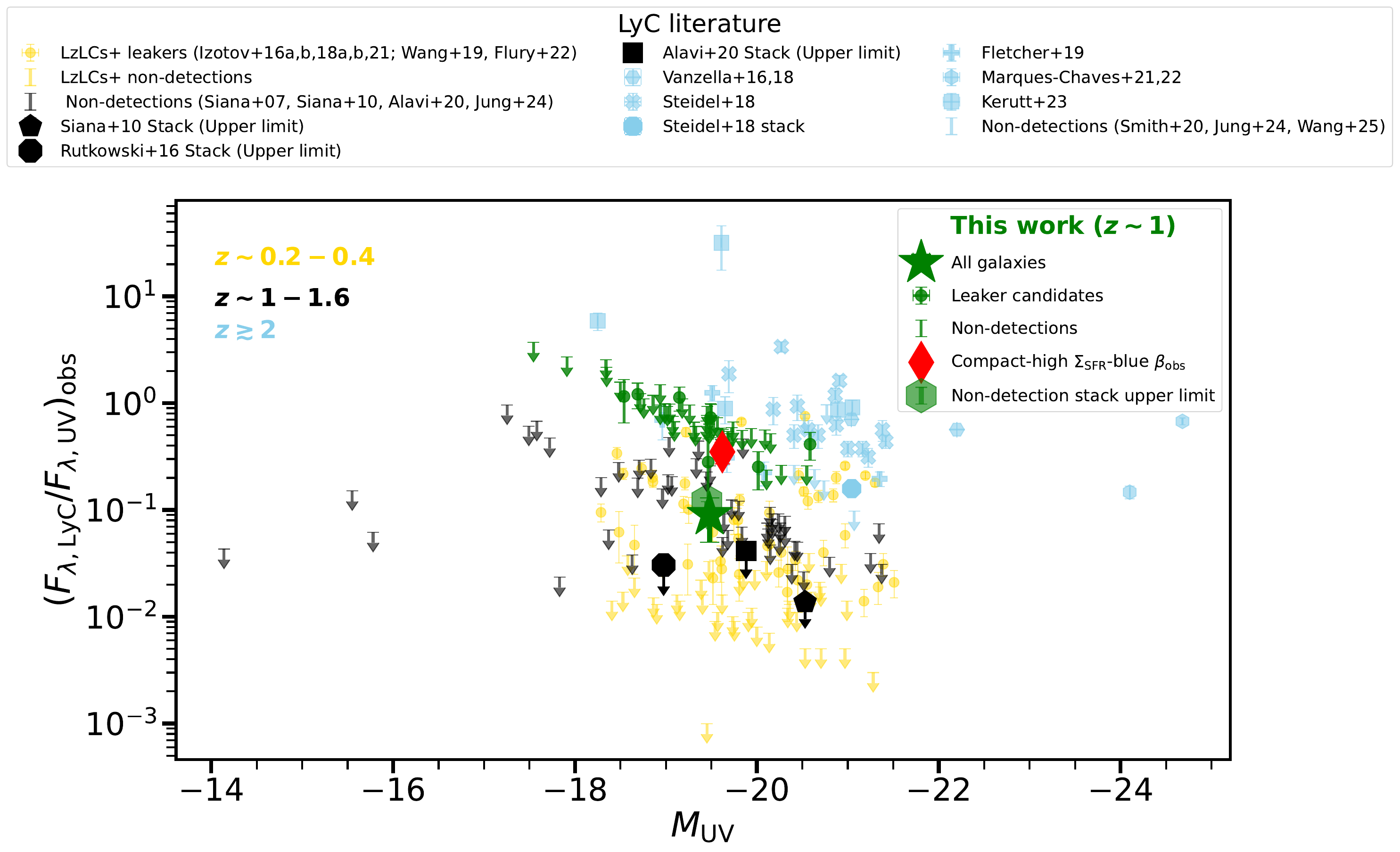} 
\caption{Absolute UV magnitude ($M_{\text{UV}}$) versus the observed LyC-to-nonionizing UV flux density ratio, corrected for IGM attenuation, for our sample (green) and for galaxies from the literature at low ($z \sim 0.2$–0.4, yellow), intermediate ($z \sim 1$–1.6, black), and high redshifts ($z \gtrsim 2$, skyblue). Both individual detections and non-detections, including upper limits from stacked samples, are shown. Upper limits correspond to $3\sigma$ in our work and in the studies of \citet{Siana07, Siana10, Rutkowski16, Alavi20}, and to $1\sigma$ in the works of \citet{Smith20, Jung24, Wang25}. The green star marks the mean ratio from the stack of all 49 galaxies in our sample, while the green hexagon denotes the 3$\sigma$ upper limit from the stack of 42 LyC non-detections. The red diamond highlights the ratio from the stack of 10 compact galaxies with high-$\Sigma_{\text{SFR}}$ and blue UV continuum slopes in our sample, as defined in Section \ref{sec:compact}.} 
\label{fig:lit_compare}
\end{figure*}

In Figure \ref{fig:lit_compare}, we present our measured LyC-to-nonionizing UV flux density ratios as a function of absolute UV magnitude. For comparison, we also show results (detections and upper limits) from other LyC studies at $z\sim0.2$–0.4 (the LzLCS+ sample \cite{Izotov2016,Izotov16b,Izotov18,Izotov18b,Izotov21,Wang19,Flury22a}), $z\sim1$–1.5 \citep{Siana07,Siana10,Rutkowski16,Jung24}, and $z\gtrsim2$ \citep{Vanzella16,Vanzella18,Steidel18,Fletcher19,Smith20,MarquesChaves21,MarquesChaves22,Kerutt23,Wang25}. For the studies at $z \gtrsim 1$, the reported flux density ratios are corrected for IGM attenuation, typically using the mean IGM transmission value adopted in each study.
When considered alongside previous studies at comparable redshifts ($z\sim1$–1.5; \citealt{Siana07,Siana10,Rutkowski16,Jung24}), the absence of a significant detection in the stack of galaxies that are individually not detected in LyC, suggests that typical star-forming galaxies at these epochs exhibit generally low LyC escape fractions. These earlier works probed diverse galaxy populations, ranging from UV-bright galaxies \citep{Siana10} to UV-faint galaxies with blue UV continuum slopes \citep{Jung24}, as well as low-mass star-forming galaxies \citep{Rutkowski16,Alavi20}, including populations whose properties overlap with confirmed LyC leakers. The high LyC detection rate of the LzLCS+ sample ($\sim56\%$) reflects its targeted selection of galaxies with properties strongly linked to LyC escape, namely, elevated O32 ratios, blue UV slopes, high $\Sigma_{\text{SFR}}$, or a combination thereof. At higher redshifts ($z\sim3$), one of the few statistical studies, the Keck Lyman Continuum Spectroscopic Survey (KLCS) conducted by \citet{Steidel18}, reports a low LyC detection fraction ($\sim12\%$), similar to what we obtain in this work ($\sim 14 \%$). Furthermore, the mean LyC-to-nonionizing UV flux density ratio derived from the spectral stack of their non-detections is lower by a factor of $\sim10$ compared to that obtained from the stacked spectra of individual LyC detections, which further indicates that the average LyC escape fraction is driven by these rare LyC leakers. It should be noted, however, that two of the strongest individual LyC detections of \cite{Steidel18} show evidence of contamination \citep{Pahl21}. Additionally, comparing detection rates across different studies should ideally take into account several factors beyond just the selection strategy, for instance, the increasing IGM optical depth to LyC at higher redshifts, the specific rest-frame LyC wavelength probed, the type of survey (spectroscopic or imaging), and the overall sensitivity of the study. Contemporary line-of-sight IGM attenuation models (e.g., \citealt{Inoue14}), which populate absorbers along the path from the source to the observer in a Poissonian manner, predict a bimodal distribution of IGM transmissions at $z\sim1$–4 \citep{Siana07,Bassett21}. One peak corresponds to a substantial fraction of sightlines with high IGM optical depths, and the prominence of this peak increases with redshift. Explicitly accounting for this distribution may provide a more accurate estimate of the true incidence rate of high–escape fraction leakers (see, for e.g., the Monte Carlo experiment of \citealt{Siana07}).

\section{Summary} 
\label{sec:summary}
In this study, we investigate LyC leakage using UVIT F154W imaging from the AUDF-south survey, focusing on 49 galaxies in the redshift range $z\sim1$–1.5 that are likely to be star-forming. The selection required galaxies to have secure redshifts, detectable H$\alpha$ emission, and minimal foreground contamination, resulting in a sample spanning a diverse range of properties. Of these, the majority (42/49) show no LyC detection, with 41 of the 42 having S/N$< 2$ in the F154W band. To constrain the average LyC escape in this epoch and to explore which galaxy properties correlate with LyC leakage, we perform stacking analyses. In the deepest regions of the F154W band, the stacked image of our entire sample reaches a depth nearly two magnitudes fainter than the individual observations.

The key results from the stacking analyses are as follows:
\begin{enumerate}
\item The stack of the 42 individual non-detections does not yield a significant signal, and only a marginal detection is obtained when all galaxies are included. This suggests that the majority of galaxies do not appreciably leak LyC photons, and the average escape is instead driven by a rare subset of leakers. 

\item We find significant detections in the bin of low-mass galaxies and marginal detections in bins of galaxies with blue UV continuum slopes, compact morphologies, or high $\Sigma_{\text{SFR}}$. In each case, however, the signal is dominated by galaxies that are already identified as LyC leakers. We find no evidence that LyC leakage depends on galaxy orientation.

\item The low-mass bin exhibits a higher observed LyC-to-nonionizing UV flux density ratio compared to high-mass galaxies. Interpreted in terms of escape fractions, this implies $\langle f_{\text{esc,abs}}\rangle^{\text{low-mass}}>\langle f_{\text{esc,abs}}\rangle^{\text{high-mass}}$, though with important caveats. Applying the same caveats, the average escape fraction across the full sample remains below a few percent ($\langle f_{\text{esc,abs}}\rangle \lesssim 0.05$).

\item Motivated by empirical results at low redshift, we isolate galaxies in our sample that are compact, have high $\Sigma_{\text{SFR}}$ and a blue UV continuum slope. While only $\sim20\%$ (10 galaxies) satisfy these criteria, the stack of this subsample shows a detection, even when restricted to galaxies individually undetected in LyC. This suggests that these properties may provide a promising basis for identifying LyC leakers in future surveys.
\end{enumerate}

Some of the data presented in this article were obtained from the Mikulski Archive for Space Telescopes (MAST) at the Space Telescope Science Institute. The specific observations analyzed can be accessed via the following \dataset[doi:10.17909/tm5e-db10]{https://doi.org/10.17909/tm5e-db10}.

\textit{Acknowledgements:} We thank Pushpak Pandey for useful discussions. We gratefully acknowledge the MUSE HUDF, MUSE WIDE, HLF, JADES, CLEAR, and 3D-HST teams for their efforts in obtaining, processing, and publicly releasing the datasets used in this study, without which this work would not have been possible. We thank the anonymous referee for their useful comments that have improved this manuscript.

\facilities{AstroSat (UVIT), HST (WFC3, ACS-WFC, WFC3-UVIS), JWST (NIRCam), VLT (MUSE)}


\software{astropy \citep{2013A&A...558A..33A,2018AJ....156..123A},  
          photutils \citep{Bradley23} 
          }

\appendix
\section{{Source information}}
{In Table \ref{tab:info_tar}, we list the source identifications along with our estimates of the absolute UV magnitude ($M_{\text{UV}}$) and the observed UV continuum slope ($\beta_{\text{obs}}$) for the 49 sources analyzed in this work. We note that these values differ from the previously published ones for the LyC leaker candidates \citep{Maulick25}, owing to the different methodology adopted here, which is based solely on photometric measurements rather than full SED fitting.}

\begin{table}[]
\centering
\caption{{Identification of sources analyzed in this work. The CLEAR, 3D-HST, MUSE HUDF, and MUSE WIDE IDs correspond to the corresponding identification numbers in the catalogs of \cite{Simons23}, \cite{Momvheva16}, \cite{Bacon23}, and \cite{Herenz17} respectively.}} \label{tab:info_tar}
\scalebox{0.92}
{\begin{tabular}{ccccccccc}
R.A. (deg) & Decl. (deg) & CLEAR ID & 3D-HST ID & MUSE HUDF ID & MUSE WIDE ID & $z$   & $M_{\text{UV}}$    & $\beta_{\text{obs}}$  \\ \hline
\multicolumn{9}{c}{\textbf{LyC non-detections}} \\ 
\hline
53.0520      & -27.7227      & 40192    & -         & -            & -            & 1.30 & $-19.93 \pm 0.07$  & $-2.35 \pm 1.32$     \\
53.1782      & -27.7831      & 28052    & 28052     & 935          & -            & 1.12 & $-19.61 \pm 0.02$  & $-1.49 \pm 0.21$     \\
53.1559      & -27.7950      & 24947    & 24947     & 934          & -            & 1.09 & $-19.48 \pm 0.02$  & $-1.51\pm0.18$       \\
53.1641      & -27.7874      & 26698    & 26698     & 30           & -            & 1.10 & $-18.99 \pm 0.03$  & $-1.83 \pm 0.39$     \\
53.1451      & -27.8099      & 21740    & 21740     & -            & 133043094    & 1.25 & $-19.18 \pm 0.04$  & $-0.90 \pm 0.22$     \\
53.1594      & -27.7751      & 29815    & 29815     & 36           & -            & 1.22 & $-18.35 \pm 0.07$  & $-1.02 \pm 0.20$     \\
53.1669      & -27.7978      & 24296    & 24296     & 978          & -            & 0.98 & $-19.09 \pm 0.02$  & $-2.06 \pm 0.16$     \\
53.1921      & -27.7872      & 26807    & 26807     & 913          & -            & 1.09 & $-19.63 \pm 0.02$  & $-1.23 \pm 0.12$     \\
53.1452      & -27.7780      & 29155    & 29155     & 924          & -            & 1.10 & $-18.72 \pm 0.04$  & $0.06 \pm 0.77$      \\
53.1038      & -27.7272      & 39253    & 39253     & -            & -            & 1.22 & $-19.26 \pm 0.12$  & $-2.64 \pm 0.70$     \\
53.1091      & -27.7225      & 40278    & 40278     & -            & -            & 1.11 & $-19.44 \pm 0.08$  & $-3.99 \pm 0.12$     \\
53.1281      & -27.7216      & 40392    & 40392     & -            & -            & 1.03 & $-20.11 \pm 0.04$  & $-1.38 \pm 0.12$     \\
53.1491      & -27.7743      & 29699    & 29699     & 928          & -            & 1.09 & $-17.54 \pm 0.11$  & $1.00 \pm 0.39$      \\
53.1554      & -27.7661      & 31529    & 31529     & 985          & -            & 1.10 & $-17.91 \pm 0.08$  & $0.14 \pm 0.54$      \\
53.1766      & -27.7855      & 27565    & 27565     & 996          & -            & 1.32 & $-18.34 \pm 0.08$  & $-0.93 \pm 1.73$     \\
53.1952      & -27.8140      & 20873    & 20873     & -            & 126060171    & 1.01 & $-18.75 \pm 0.04$  & $-2.55 \pm 0.40$     \\
53.1441      & -27.8195      & 19759    & 19759     & -            & 133041092    & 1.04 & $-18.94 \pm 0.03$  & $-2.40 \pm 0.25$     \\
53.1357      & -27.7978      & 24108    & 24108     & 1091         & -            & 1.31 & $-19.45 \pm 0.03$  & $-2.32 \pm 0.37$     \\
53.1291      & -27.7546      & 33930    & -         & -            & 140044111    & 1.08 & $-19.69 \pm 0.07$  & $-2.47 \pm 1.55$     \\
53.1873      & -27.8349      & -        & 16740     & -            & 119043082    & 1.30 & $-19.74 \pm 0.02$  & $-1.70 \pm 0.15$     \\
53.1619      & -27.8459      & -        & 13920     & -            & 118031091    & 1.02 & -                  & $0.31 \pm 0.42$      \\
53.1485      & -27.8407      & -        & 15291     & -            & 117030079    & 1.13 & -                  & $-2.79 \pm 0.24$     \\
53.1085      & -27.8633      & -        & 10102     & -            & 110032078    & 1.24 & $-19.72 \pm 0.02$  & $-1.79 \pm 0.19$     \\
53.1273      & -27.8387      & -        & 15831     & -            & 122044143    & 1.00 & $-19.32 \pm 0.02$  & $-1.69 \pm 0.26$     \\
53.0899      & -27.8283      & -        & 18022     & -            & 108032158    & 1.11 & $-19.02 \pm 0.03 $ & $-1.62 \pm 0.42$     \\
53.0834      & -27.8253      & -        & 18676     & -            & 107038155    & 1.02 & $-19.08 \pm 0.02$  & $-2.54 \pm 0.10$     \\
53.0956      & -27.8220      & -        & 19304     & -            & 107007085    & 1.30 & $-18.94 \pm 0.05$  & $-3.19\pm 0.35$      \\
53.0854      & -27.8082      & -        & 21822     & -            & 106045099    & 1.11 & $-19.68 \pm 0.02$  & $-1.11 \pm 0.28$     \\
53.0829      & -27.8062      & -        & 22375     & -            & 106046100    & 1.11 & $-19.04 \pm 0.03$  & $-2.63\pm 0.47$      \\
53.1092      & -27.7953      & -        & 24623     & -            & 130046072    & 1.41 & $-19.50 \pm 0.03$  & $-2.63 \pm 0.13$     \\
53.1096      & -27.7883      & -        & 26599     & -            & 130032058    & 0.99 & $-20.27 \pm 0.02$  & $-1.70 \pm 0.18$     \\
53.1158      & -27.7877      & -        & 26452     & -            & 130037063    & 1.19 & $-19.45 \pm 0.02$  & $-2.99 \pm 0.19$     \\
53.1009      & -27.7834      & -        & 27698     & -            & 131047151    & 1.08 & $-18.50 \pm 0.05$  & $-2.66 \pm 0.22$     \\
53.0776      & -27.7848      & -        & 27599     & -            & 143056142    & 1.38 & $-20.01 \pm 0.02$  & $-1.94 \pm 0.06$     \\
53.0686      & -27.7840      & -        & 27632     & -            & 143051136    & 1.22 & $-19.84 \pm 0.03$  & $-1.87 \pm 0.03$     \\
53.0483      & -27.7731      & -        & 30280     & -            & 146067352    & 1.22 & $-20.55 \pm 0.01$  & $-1.43 \pm 0.19$     \\
53.0761      & -27.7812      & -        & 28288     & -            & 143052137    & 1.23 & $-19.56 \pm 0.02$  & $-2.53 \pm 0.42$     \\
53.0884      & -27.7740      & -        & 30129     & -            & 142046161    & 1.22 & $-19.52 \pm 0.03$  & $-2.21 \pm 0.94$     \\
53.1394      & -27.7674      & 31444    & 31444     & -            & 139067324    & 1.29 & -                  & $-1.58 \pm  0.22$    \\
53.1116      & -27.7585      & -        & 33202     & -            & 137072148    & 1.12 & $-18.86 \pm 0.04$  & $-2.67 \pm 0.60$     \\
53.1360      & -27.8136      & 20915    & 20915     & -            & 133046097    & 1.34 & $-20.15 \pm 0.02$  & $-1.90 \pm 0.02$     \\
53.1013      & -27.7728      & -        & 30384     & -            & 131044148    & 1.02 & $-19.31 \pm 0.02$             & $-2.43 \pm 0.45$     \\ \hline
\multicolumn{9}{c}{\textbf{LyC leaker candidates}} \\ 
\hline
53.0559      & -27.7212      & 40553    & 40553     & -            & -            & 1.04     & $-19.15 \pm 0.10 $ & $-2.83 \pm 1.06$     \\
53.1713      & -27.7930      & 25304    & 25304     & 1002         & -            & 0.99     & $-18.69 \pm 0.03$  & $-1.68 \pm 0.52$     \\
53.1128      & -27.7200      & 40878    & 40878     & -            & -            & 1.11     & $-20.02 \pm 0.05$  & $-1.94 \pm 0.50$     \\
53.1281      & -27.7293      & 38849    & 38849     & -            & -            & 1.42     & $-20.58 \pm 0.05$  & $-2.23 \pm 0.40$     \\
53.1588      & -27.7706      & 30520    & 30520     & 13           & -            & 1.00     & $-19.46 \pm 0.02$  & $-1.04 \pm 0.24$     \\
53.1659      & -27.7816      & 28448    & 28448     & 16           & -            & 1.10     & $-18.54 \pm 0.05$  & $-0.44 \pm 0.03$     \\
53.0850      & -27.7783      & -        & 29027     & -            & 141050163    & 1.23     & $-19.49 \pm 0.02$  & $-3.22 \pm 0.75$    
\end{tabular}}
\end{table}



\bibliography{stack_z1_to_arxiv_August25}{}
\bibliographystyle{aasjournal}



\end{document}